\documentclass[journal]{IEEEtran}

\usepackage{graphicx}

\usepackage[cmex10]{amsmath}
\usepackage{latexsym}

\usepackage{textcomp}

\usepackage{microtype} 

\usepackage{float} 

\usepackage{hyperref} 

\usepackage{lettrine} 
\usepackage{paralist} 

\begin{document}

\title{\vspace{0mm}\fontsize{22pt}{0pt}\selectfont\textbf{Classic Sliding Mode Control \\ From First Principles \\ \fontsize{12pt}{0pt}\textnormal{Version 2.1b (28/04/2017)}}}

\author{
	\textsc{Ogbeide Imahe}{} 
	\\[0.5mm] 
	\thanks{Author emails: imaheo@abuad.edu.ng}  \thanks{ogbeide.imahe@alumni.manchester.ac.uk}}

\markboth{}%
{O. Imahe\MakeLowercase{\textit{}}}
%



\maketitle 


\begin{abstract} 
This is an intuitive introduction to classic sliding mode control that shows how the associated assumptions and condition for its use arise in the context of a derivation of the method. It derives a controller that obviates the need for the assumption of any sign for the control input vector, answers why it is said that it deals only with matched disturbances and why a system that it may apply to \textquotesingle must\textquotesingle \hspace*{1pt} be linear in the control signal. Additionally, it may be viewed as an example of how a control design method might be developed, adding to its pedagogical usefulness.

\end{abstract}





\section{Introduction}
The sliding mode control design method is suited to a large group of systems, both linear and nonlinear; including electromechanical, power, robotics, and aerospace systems; where matched disturbances are a significant type of disturbance to be dealt with. Additionally, the resulting control signal may be relatively easy to compute and implement. \cite{SMCinElectroMechs:Utkin} \cite{SMC_TheoryAppl:EdwardsSpurgeon} \cite{IntuitSMC:Shtessel} It is therefore a useful tool in the control engineer's toolbox, and may thus be considered important for students of control to know.

This note advances a coherent and intuitive stream of thought that leads to the development of the sliding mode control design method; explaining classic sliding mode control design in the context of how its characteristics and application conditions may be logically arrived at from first principles. Thus, in addition to serving as an introductory literature on the subject, it may also be seen as an example of deriving controller design methods. This approach is different from, but complementary to, various other treatments, for instance, in \cite{SMCinElectroMechs:Utkin}, \cite{SMC_TheoryAppl:EdwardsSpurgeon}, \cite{IntuitSMC:Shtessel} that focus on describing the method, practical performance characteristics, advancements, and usage; the note will not cover these areas.

Single-input single-output (SISO) systems are considered. And the coverage is general and theoretical. A basic knowledge of control theory and associated mathematics is helpful to follow the discussion. And Lyapunov stability theory can be gleaned from the coverage, so it is not a necessary prerequisite.

The next section presents the derivation, after which follows the concluding section.

\section{Classic Sliding Mode Control \label{s:CSMC}}
Consider a SISO system and its output equation in the abstract:
\begin{equation}
\label{eq:generalSystem}
\begin{array}{l}
\dot{x} = f(x) + g(x, u) + g(x,d) + w_u \\
y = s(x) \\
\end{array},
\end{equation}
where the output $ y $, the control input $ u $, and a disturbance input $ d $ are scalar. The functions $ f(\cdot) $ and $ g(\cdot)  $ have the same row dimension as the dimension of the system state vector $ x $. Disturbance input $ d $ acts through the same channel as the control input, and is thus matched with it; disturbance vector $ w_u $ has elements that do not feed into the system through the same channel as a control input, and are thus unmatched with any control signal. 

The objective is to find a generally applicable controller design paradigm (hence the abstract view of the system under investigation) that results in controllers that stabilise the system output. A controller that comes from following this paradigm should therefore drive the output of the system to a defined equilibrium point or desired value. And Lyapunov stability theory provides an avenue to derive such a controller.

Therefore, let us take a candidate Lyapunov function based on the output as
\begin{equation}
\label{eq:Lyap_func_quadratic}	
V = ps^2(x), \qquad p > 0.
\end{equation}
(The quadratic Lyapunov function is a common choice.) The system will behave stably if $ \dot{V} < 0 $ when $ V \neq 0 $, $ \dot{V} = 0 $ when $ V = 0 $, and $ V(x) $ tends to increase or decrease as the magnitude of $ x $ does. Hence, upon the action of the controller, the value of $ V $ must tend to zero, and in a finite time arrive within a very small region of it so that it may be approximated as zero---ideally reaching zero. 



Where $ V = 0 $ (implies that $ s = 0 $) defines the objective or equilibrium point, we can specify a suitable stable first-order dynamics in $ V $ to achieve the goal:
\begin{equation}
\label{eq:vdotLinearSys}
\dot{V} = -nV^{0.5}, \qquad  n > 0 .
\end{equation}
The above dynamics ensures that $ V $ goes to zero, and reaches it in finite time \cite{IntuitSMC:Shtessel}. As $ V $ tends to zero, $ s(x) = s = s(x(t)) $ also tends to zero, thereby satisfying the objective. 

Since
\begin{equation*}
	\dot{V} = 2ps\dot{s} ,
\end{equation*}
we have
\begin{equation*}
	2ps\dot{s} = -nV^{0.5} .
\end{equation*}
The RHS (right-hand-side) of which must be negative in order to satisfy stability requirements (Lyapunov stability theory). Thus choose  $ V^{0.5} = p^{0.5} |s| $ to guarantee this, thereby giving
\begin{equation*}
2ps\dot{s} = -n p^{0.5} |s| .
\end{equation*}
Without loss of generality, let $ p = 0.25 $ to eliminate the number $ 2 $ on the left-hand-side (LHS) above, and to thus yield the more general looking equation,
\begin{equation}
\label{eq:hHdotIsMinusNabsH}
s\dot{s} = -n |s| .
\end{equation} 

Dividing both sides of \eqref{eq:hHdotIsMinusNabsH} by $ s $ yields,
\begin{equation}
\label{eq:DEwithDiscRHS}
 \dot{s} = -n \cdot sgn(s) .
\end{equation}
which is structurally a rephrasing of Lyapunov stability theory, with the appropriate specification on $ V $. It is also a differential equation with a discontinuous RHS, explored in \cite{DEwithDiscontinuousRHS:Filippov} and \cite{SlidingModesInCtrlAndOptim:Utkin}.

To solve \eqref{eq:DEwithDiscRHS} simply, let us consider the definition of the signum function ($ sgn(\cdot) $) as defining three cases, and then solve for $ s(x(t)) $ in each case.

\paragraph*{CASE 1}
$ s>0 $, therefore $ sgn(s) = 1 $ and $ \dot{s} = -n  $, yields,
\begin{equation*}
\int_{s(x(0))}^{s(x(t))} \mathrm{d}s = -n \int_{0}^{t} \mathrm{d}t ,
\end{equation*}
which results in
\begin{equation*}
s(x(t)) - s(x(0)) = -nt ,
\end{equation*}
and thus 
\begin{equation*}
s(x(t)) = -nt + s(x(0)) ,
\end{equation*}
which is the equation of a straight line.

Because $ s>0 $, $ s(x(t)) $ is a reducing value, decreasing by $ nt $ and tending towards zero, it will reach zero at a certain time $ t = t_r $. This leads us to the next case.

\paragraph*{CASE 2} $ s=0 $, therefore $ sgn(s) = 0 $ and $ \dot{s} = 0  $.
Integrating, as above, yields $ s(x(t)) - s(x(0)) = 0 $. Hence $ s $ remains the same; if it is zero, it remains zero.

\paragraph*{CASE 3} $ s<0 $, therefore $ sgn(s) = -1 $ and $ \dot{s} = n  $.
After the appropriate integration of both sides,
\begin{equation*}
s(x(t)) - s(x(0)) = nt ,
\end{equation*}
and thus 
\begin{equation*}
s(x(t)) = nt + s(x(0)) ,
\end{equation*}
which is also the equation of a straight line.

Because $ s<0 $, $ s(x(t)) $ is an increasing value, increasing by $ nt $, and thus tending towards zero. It will reach zero at a certain time $ t = t_r $.

The solution equations from the above three cases can be combined to form a single equation solution:
\begin{equation*}
s(x(t)) = -nt \cdot sgn \left( s(x(0)) \right) + s(x(0)) .
\end{equation*}
From which the time $ t_r $ that it takes for $s(x(0)) $ to reach $ s(x(t)) = 0 $, is computed to be,
\begin{equation*}
t_r = n^{-1} |s(x(0))| .
\end{equation*}
This is the convergence time. Let us call it the \emph{reaching time}, considering that it is the time taken to reach $ s(x)=0 $. When $ s(x) \neq 0 $, the system is made to approach the sliding mode and is thus said to be in the \emph{reaching phase} or \emph{reaching mode}. We call \eqref{eq:DEwithDiscRHS} a \emph{reaching law} or \emph{reaching condition} since its form permits the design of controllers that ensure that the sliding mode is reached. 

Hence, regardless of the initial value of $ s $, it should, by the action of the control signal determined from \eqref{eq:hHdotIsMinusNabsH} or \eqref{eq:DEwithDiscRHS}, reach $ s=0 $ after a time $ t_r $ and remain there afterwards. (This proof of the finite-time convergence of $ s $ to zero given \eqref{eq:DEwithDiscRHS} has thus also shown the finite-time convergence of $ V $ given \eqref{eq:vdotLinearSys}.

If we take the output $ s(x) $ to be $ x_1 $ or some function of it, Lyapunov stability analysis would not show that the other states behave stably. To remove this blind spot, let $ s $ be a function of all the states. This then implies that $ s(x(t))=0 $ must represent a stable trajectory of the states of the system so that the control objective is achieved. Thus  $ s = 0 $ encapsulates a stable system.

We call $ s(x)=0 $ the \emph{sliding mode}, and the evolution of the states on it defines the \emph{sliding surface}. Intuitively, $ s(x)=0 $ is designed to respect the relationship \eqref{eq:generalSystem} in such a way that it is a stable evolution of the trajectories of the system; its choice cannot be arbitrary. 

The rate of change of each state of the system model (\ref{eq:generalSystem}) is generally a function of other states. If the sliding mode is designed using only the states of the system, and such that the state derivative that admits the input and matched disturbance does not reflect in the formulation, the resulting dynamics in the sliding mode appears invariant to these inputs. This is a significant advantage.

However, by the definition of unmatched disturbance, their effects will remain since the remaining state derivatives are implicit in the states. Hence this approach would not deal with unmatched disturbances unless they are exactly known, perhaps, so that they can be exactly compensated for.

For sure, $ s(x) $ would yield the control input variable on its first derivative so that (\ref{eq:DEwithDiscRHS}) can be used for controller determination. This makes the dynamics of the sliding mode, one order less than that of the model. 

Let us now rewrite \eqref{eq:hHdotIsMinusNabsH} as
\begin{equation*}
\label{eq:hdotexp}
 s\dfrac{\partial{s}}{\partial{x}} \dot{x} = -n|s| ,
\end{equation*}
and then expand it to
\begin{equation*}
 s\dfrac{\partial{s}}{\partial{x}} (f(x) + g(x,u) + g(x,d) + w_u) = -n |s| .
\end{equation*}
To make $ u $ directly determinable, let the applicable set of systems be such that $ g(x,u) = b(x)u = bu $: systems that are linear in the control variable. This also makes $ g(x,d) = b(x)d = bd $, therefore
\begin{equation}
\label{eq:hhdotExpStabilityEq}
s\dfrac{\partial{s}}{\partial{x}} (f + bu + bd + w_u) = -n |s|
\end{equation}
($ f(x) = f $ also, for notational convenience). This makes
\begin{equation}
\label{eq:uWithUnmatched}
u = - (\dfrac{\partial{s}}{\partial{x}} b )^{-1} \left( n (sgn(s) + (\dfrac{\partial{s}}{\partial{x}} f ) +  (\dfrac{\partial{s}}{\partial{x}} w_u) \right) - d .
\end{equation}

This says that $ w_u $ contributes to a perception of additional input disturbance quantified by $ -(\frac{\partial{s}}{\partial{x}} b )^{-1} (\frac{\partial{s}}{\partial{x}} w_u)  $, and would need to be considered in the control design. Additionally, notice that a sign switch occurs when $ s $ crosses zero; so that $ s $ is also known as the \emph{switching function}. Since the disturbance added to control $ u $ due to $ w_u $ is a function of $ s $, there might be a trade-off relationship between the permissible limits of the control signal and the switching function chosen.

Assuming that $ w_u $ is known and $ d $ is unknown, $ d $ is replaced with a guess $ d_g $ so that the control signal is determinable. Thus we have
\begin{equation*}
\label{eq:uWithd_g}
u = - (\dfrac{\partial{s}}{\partial{x}} b )^{-1} \left( n (sgn(s) + (\dfrac{\partial{s}}{\partial{x}} f ) +  (\dfrac{\partial{s}}{\partial{x}} w_u) \right) - d_g .
\end{equation*}

\noindent Substituting the above equation for $ u $ in  \eqref{eq:hhdotExpStabilityEq} yields
\begin{equation*}
-n|s| - s\dfrac{\partial{s}}{\partial{x}} b(d_g - d)  = - n|s| - m, \qquad m>0 .
\end{equation*}
The RHS has been adjusted by $ -m $ to make the equation balanced and to satisfy the requirements for stability. If $ d $ were known, then $ d_g $ could be chosen as equal to it so that the corresponding term on the LHS becomes zero, making $  m = 0 $ on the RHS also.

Controller parameter $ d_g $ may be derived from the equation above using corresponding left-- and right--hand--sides:
\begin{equation*}
-s\dfrac{\partial{s}}{\partial{x}} b(d_g - d) = -m .
\end{equation*}
To eliminate the need to use the magnitude of $ s\frac{\partial{s}}{\partial{x}} b $ in determining $ d_g $, and thereby simplify the above equation, choose $ m = |s\frac{\partial{s}}{\partial{x}}b|m_m \geq 0 $, where $ m_m $ is some positive number. The above equation could then be written as,
\begin{equation*}
- sgn ( s\dfrac{\partial{s}}{\partial{x}} b) (d_g - d) = -m_m .
\end{equation*}
If $ sgn ( s\frac{\partial{s}}{\partial{x}} b ) = 1 $, make $ d_g > 0 $ and $ d_g > d $. And if $ sgn ( s\frac{\partial{s}}{\partial{x}} b ) = -1 $, make $ d_g < 0 $ and $ |d_g| > |d| $. Therefore, choose $ d_g = d_m sgn ( s\frac{\partial{s}}{\partial{x}} b ) $, where $ d_m $ is an estimate of the maximum absolute value of $ d $, therefore making it a specification for the controlled system.

Likewise, with a known $ d $, and for $ w_u $ that is unknown, we replace $ (\frac{\partial{s}}{\partial{x}} b )^{-1} (\frac{\partial{s}}{\partial{x}} w_u) = w_{ui} $ in (\ref{eq:uWithUnmatched}) with $ w_{uig} $ to get the control equation,
\begin{equation*}
\label{eq:uWith_wuig_d}
u = - (\dfrac{\partial{s}}{\partial{x}} b )^{-1} \left( n sgn(s) + (\dfrac{\partial{s}}{\partial{x}} f ) \right) - w_{uig} - d .
\end{equation*}

Hence, with the same reasoning used to determine $ d_g $, we set $ w_{uig} = w_{uim} sgn(s) sgn ( s\frac{\partial{s}}{\partial{x}} b ) $, where $ w_{uim} $ is an estimated maximum absolute value for $ w_{ui} $.

With the estimates of $ d_m $ and $ w_{uim} $, the control signal is now written as
\begin{equation*}
\begin{array}{l}
u = - (\dfrac{\partial{s}}{\partial{x}} b )^{-1} \left( n sgn(s) + (\dfrac{\partial{s}}{\partial{x}} f ) \right) - w_{uig} - d_g \\
d_g = d_m sgn(s) sgn ( \dfrac{\partial{s}}{\partial{x}} b ) \\
w_{uig} = w_{uim} sgn(s) sgn ( \dfrac{\partial{s}}{\partial{x}} b ) , \\
\end{array}
\end{equation*}
and then rearranged to
\begin{equation}
\label{eq:finalControlSignal}
\begin{array}{ll}
u =  - (\dfrac{\partial{s}}{\partial{x}} b )^{-1} (\dfrac{\partial{s}}{\partial{x}} f ) &  \\
-  \{ (\dfrac{\partial{s}}{\partial{x}} b )^{-1} [n + \frac{\partial{s}}{\partial{x}} w_{um} sgn ( \dfrac{\partial{s}}{\partial{x}} b )] + d_m sgn ( \dfrac{\partial{s}}{\partial{x}} b ) \} sgn(s) , &\\
\end{array}
\end{equation}
with $ w_{um} $ the estimated $ w_{u} $ vector that produces $ w_{uim} $.
We note also that $ (\frac{\partial{s}}{\partial{x}}b) $ must be invertible. 

The stability equation \eqref{eq:hhdotExpStabilityEq} becomes 
\begin{equation*}
s\dfrac{\partial{s}}{\partial{x}} ( f + bu + bd + w_u ) = -n |s| - (m_m |s| + m_u |s|) \left| ( \dfrac{\partial{s}}{\partial{x}} b ) \right|,
\end{equation*}
with a new RHS, and where $ m_u $ is the analog of $ m_m $ with respect estimating $ w_{um} $. Also $ \eqref{eq:hHdotIsMinusNabsH} $ becomes
\begin{equation*}
s\dot{s} = - \left( n + (m_m |s| + m_u |s|) \left| ( \dfrac{\partial{s}}{\partial{x}} b ) \right| \right) |s| ,
\end{equation*}
leading to the differential equation
\begin{equation*}
\dot{s} = - \left( n + (m_m |s| + m_u |s|) \left| ( \dfrac{\partial{s}}{\partial{x}} b ) \right| \right) sgn(s),
\end{equation*}
where $ \frac{\partial{s}}{\partial{x}} b $ is bounded, and $ m_m \geq 0 $ is a function of $ d_g - d $ which, although uncertain because of $ d $, is positive for the right estimate of $ d_m $. The reaching time $ t_r $ is also uncertain because it depends on the disturbance signal $ d $, but it remains bounded for the right estimate of $ d_m $:
\begin{equation*}
t_r = \left( n + (m_m |s| + m_u |s|) \left| ( \dfrac{\partial{s}}{\partial{x}} b ) \right| \right)^{-1} |s(x(0))|.
\end{equation*}

\section{Conclusion}
This note has presented sliding mode control using the approach of a derivation; which, presumably, gives it added pedagogically usefulness. Additionally it may be viewed as an example of how to develop a control design method and as material to encourage exploration and innovation.

Characteristic of sliding mode control is the use of a controller based on a switching function that encapsulates a stable dynamical system. For the requirement that a suitable switching function be found, this is feasible for a large class of systems \cite{SMCinElectroMechs:Utkin}, \cite{SMC_TheoryAppl:EdwardsSpurgeon}, \cite{IntuitSMC:Shtessel}, \cite{SlidingModesInCtrlAndOptim:Utkin}. Another characteristic is the invariance to matched disturbance, and an impotence against unmatched disturbance without some modification or augmentation of the method. Unmatched disturbance also appears as some additional perturbation through the control input channel.

Some conditions for using this controller design method are simplifying assumptions that may be ignored where appropriate or feasible. Because it is clear that it is not necessary that the system be linear in the control signal, although this assumption makes control computation easier. Also it can be seen in (\ref{eq:finalControlSignal}) that the elements of the input matrix could be negative: the typical assumption is that its elements are non-negative.

Finally, this has been an introduction to the classic method of sliding mode control, and several advanced developments and modifications can be found in the relevant references earlier given. 





\begin{thebibliography}{99} 

\bibitem{SMCinElectroMechs:Utkin}
V.~Utkin et al., Sliding mode control in electro-mechanical systems, 2nd ed. Philadelphia, PA: CRC/Taylor and Francis, 2009.

\bibitem{SMC_TheoryAppl:EdwardsSpurgeon}
C.~Edwards and S. K.~Spurgeon, Sliding Mode Control: Theory and Applications, London: Taylor and Francis, 1998.

\bibitem{IntuitSMC:Shtessel}
Y. Shtessel et al., "Intuitive Theory of Sliding Mode Control," in \textit{Sliding Mode Control and Observation}, Basel: Birkh\"{a}user, 2014, pp. 1-42.

\bibitem{DEwithDiscontinuousRHS:Filippov}
A. Filippov, Differential Equations with Discontinuous Right-Hand Sides, Kluwer Academic Publishers, 1988.

\bibitem{SlidingModesInCtrlAndOptim:Utkin}
V. I. Utkin, Sliding Modes in Control and Optimisation, Berlin: Spinger-Verlag, 1992.

%

\end{thebibliography}
\end{document}